\documentclass[%
 aip,
 jmp,%
 amsmath,amssymb,
preprint,%
]{revtex4-1}

\usepackage{graphicx}
\usepackage{dcolumn}
\usepackage{bm}

\begin{document}

\preprint{APS/}
\title{A fast model based on muffin-tin approximation to study charge transfer effects in time-dependent quantum transport simulations: doped Si-SiO$_{2}$ quantum-dot systems }

\author{I-Lin Ho}
\email{sunta.ho@msa.hinet.net}
\affiliation{ChiMei Visual Technology Corporation, Tainan 741, Taiwan, R.O.C.}

\date{\today}

\begin{abstract}
In order to quickly study quantum devices in transient problems, this work demonstrates an analytical algorithm to solve the Hartree potential associated with charge fluctuations in the time-dependent non-equilibrium green function (TDNEGF) method.
We implement the calculations in the heterojunction system of gold metals and silicon quantum dots for applications of photoelectric semiconductors in the future. Numerical results for the transient solutions are shown to be valid by comparing with the steady solutions calculated by the standard time-independent density functional method.
\end{abstract}

\maketitle
\section{Introduction}
Photoelectric bioengineering - the use of photoelectric semiconductors as functional entities in biological systems - is heralded as an alternative option for signaling communications between organisms and physical devices in future biomedicines. In particular, research on quantum dots \cite{bio1} has already revealed a variety of biologically-oriented applications, e.g. drug discovery \cite{bio2,bio3}, disease detection \cite{bio4,bio5}, protein tracking \cite{bio6,bio7}, and intracellular reporting \cite{bio8,bio9}.
A qualitative understanding of these complex processes has been accessed by perturbative electron-photon interactions associated with strong electron correlations \cite{qd_tb2}, but the quantitative agreement between the first-principles theory and experiments is still unsatisfactory from the perspective of the ground-state density functional theory (DFT) \cite{qd_tb3,steady1}.

In recent years, the majority of studies for quantum-dot electronics have focused on the time-dependent density functional theory (TDDFT) \cite{qd_tb4} that provides a more rigorous theoretical foundation \cite{tddft1}. The formalism may also be easily extended to cover the interaction of electrons with light or under environments in open quantum systems by the time-dependent non-equilibrium green function (TDNEGF) technique \cite{thesis1,TDNEGF1}, e.g. for the photon-assisted transport and fluorescence of contacted atomic devices.

Several scenarios of open quantum systems implemented with TDDFT have so far been suggested, including the ring-topology of the electronic circuit \cite{tddft2,tddft3} and the approximately-isolated atomic device \cite{tddft4,tddft5}. This present work adopts more general set-ups to study systems composed of functional atomic devices and environmental clusters as shown in Figure \ref{fig_TDNEGF}. Here, the device in the central region focuses on the Si-SiO$_{2}$ core-shell quantum dot due to its wide application spectrum. Moreover, the energetically-favored phosphorus impurities are considered for low-voltage operations. Nuria \cite{QD1} reported relevant analyses of doped Si-SiO$_{2}$ quantum dots in detail. The effect of the neighboring Au(111) electrodes upon devices is also exactly accounted for through properly defined self-energies. For numerical treatments, the initial Kohn-Sham (KS) single-particle Hamiltonians and the overlap matrices for the Si-SiO$_{2}$ quantum dots and Au electrodes are obtained by DFT calculations in SIESTA programs \cite{siesta1,siesta2}. On the basis of the holographic electron density theorem and Runge-Gross theorem, the time-dependent electron dynamics are then determined by solving equations of motion for the devices using the TDNEGF technique \cite{TDNEGF1,TDNEGF2} in own-implemented fortran programs \cite{codeF}.

Computations by TDNEGF of realistic devices having a large amount of atoms are numerically demanding, because all electron motions have to be fully resolved, leading to considerable degrees of freedom in the sub-fs time resolution. To arrive at a computationally efficient but still predictive method, this study follows the work of Chen \cite{TDNEGF1} for open quantum systems driven by time-dependent bias potentials. Furthermore, to consider the effects of the charge variations through devices, this work demonstrates a capacitive network model \cite{QC1} as an analytical Poisson solution in the muffin-tin approximation. Rather than the numerically-demanding iterative Poisson-equation solution using discretized spatial grids \cite{thesis1}, the model can quickly solve the Hartree potential associated with the charge fluctuations in the time-dependent non-equilibrium green function (TDNEGF) formulae. Numerical results are shown to be valid by comparing with the steady solutions calculated by the standard time-independent density functional method using SIESTA.

This paper is organized as follows. Section \textrm{II} describes the employed computational algorithms. Section \textrm{III} demonstrates the density of state of the gold electrode for validating the wide-band limit approximation, and depicts the characteristics of energy spectrums for quantum-dot devices. We then compute the electronic transmission functions of the quantum device coupled to the semi-infinite electrodes by the SIESTA program so as to study the steady transport dynamics. The properties of the transient electronic transport for the open quantum system are simulated by using the a fortran program, and are shown to be good versus the steady solutions in a long time limit.
Finally, Section \textrm{IV} presents concluding remarks. The mathematical derivations and relevant physical approximations for the time-dependent non-equilibrium green function (TDNEGF) formulae are stated in the appendix.

\section{Time-dependent non-equilibrium green function for quasi-one-dimensional open quantum systems}

\begin{figure*}[ht]
\includegraphics[scale=0.16]{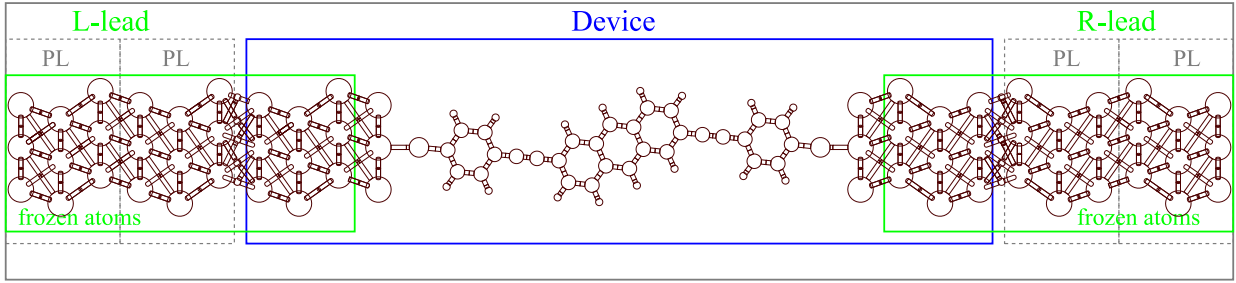}
\caption{ Schematic representation of general simulation setups for open quantum systems including Au(111) electrodes and atomic devices. }
\label{fig_TDNEGF}
\end{figure*}
Figure \ref{fig_TDNEGF} shows a regular open quantum system, including semi-infinite electrodes and atomistic devices.
The system is partitioned by several electronically-functional areas, named as L-electrode (L), device (D), and R-electrode (R).
The equation of motion (EOM) for electrons can be described by quantum dynamics:
\begin{equation}
i\mathbf{\dot{\sigma}}\left( t\right) =\left[ \mathbf{h}\left( t\right) ,
\mathbf{\sigma }\left( t\right) \right]  \label{a1}
\end{equation}
where $\mathbf{h}\left( t\right) $ is the Kohn-Sham Hamiltonian matrix, and
the square bracket on the right-hand side (RHS) denotes a commutator. The
matrix element of the single-electron density $\mathbf{\sigma }$ is defined
by $\sigma _{ij}\left( t\right) =\left\langle a_{j}^{\dagger
}(t)a_{i}(t)\right\rangle $, where $a_{j}^{\dagger }(t)$ and $a_{i}(t)$ are
the creation and annihilation operators for atomic orbitals $j$ and $i$ at
time $t$, respectively. On the basis of the atomic orbital sets for
electrons, the matrix representation of $\mathbf{\sigma }$ and $\mathbf{h}$
can be written as
\begin{equation}
\mathbf{h=}\left[
\begin{array}{cccc}
\mathbf{h}_{L} & \mathbf{h}_{LD} & 0 \\
\mathbf{h}_{DL} & \mathbf{h}_{D} & \mathbf{h}_{DR} \\
0 & \mathbf{h}_{RD} & \mathbf{h}_{R}
\end{array}
\right] ,\text{ \ }\mathbf{\sigma =}\left[
\begin{array}{cccc}
 \mathbf{\sigma }_{L} & \mathbf{\sigma }_{LD} &
\mathbf{\sigma }_{LR} \\
 \mathbf{\sigma }_{DL} & \mathbf{\sigma }_{D} &
\mathbf{\sigma }_{DR} \\
 \mathbf{\sigma }_{RL} & \mathbf{\sigma }_{RD} &
\mathbf{\sigma }_{R}
\end{array}
\right]  \label{a2}
\end{equation}
Here, $\mathbf{m}_{L}$, $\mathbf{m}_{D}$, and $\mathbf{m}%
_{R}$ ($\mathbf{m}\in \left\{ \mathbf{h},\mathbf{\sigma }\right\} $)
represent the matrix blocks corresponding to left-electrode $L$, device $D$, and right-electrode $R$
partitions, respectively. Moreover, $\mathbf{h}_{LR}$ and $\mathbf{h}_{RL}$ are
ignored due to the distant separation between L and R electrodes in common
applications. We note that the holographic electron density
theorem and Runge-Gross theorem are adopted for time-dependent electron dynamics \cite{TDNEGF1,TDNEGF2}, stating that the initial ground-state density
of the subsystem $\mathbf{\sigma }_{D}\left( t_{0}\right) $ can determine
all physical properties of systems at any time $t$. Hence $\mathbf{h}$ and $\mathbf{\sigma }$ can be approximately expressed as
functions of $\mathbf{\sigma }_{D}\left( t\right) $ for a formally
closed-form equation of motion as described below.

Placing Eq. (\ref{a2}) into Eq. (\ref{a1}), we can write the equation of motion for $
\mathbf{\sigma }_{D}$ as
\begin{eqnarray}
i\dot{\sigma}_{D,mn} &=&\sum_{\ell \in D}\left( h_{D,m\ell }\sigma _{D,\ell
n}-\sigma _{D,m\ell }h_{D,\ell n}\right) -i\sum_{\alpha =L,R,N}Q_{\alpha ,mn}
\label{a3} \\
Q_{\alpha ,mn} &\equiv &i\sum_{k_{\alpha }\in \alpha }\left( h_{D\alpha
,mk_{\alpha }}\sigma _{\alpha D,k_{\alpha }n}-\sigma _{D\alpha ,mk_{\alpha
}}h_{\alpha D,k_{\alpha }n}\right)  \label{a4}
\end{eqnarray}
Here, $m$ and $n$ denote the atomic orbital in partition D, $k_{\alpha }$
denotes the state of $\alpha $ ($\alpha $=L, R) electrode, and $Q_{\alpha }$ is the
dissipation term due to the contacts of the device with electrodes L and R. The transient current through an electrode's
interfaces can be calculated by:
\begin{eqnarray}
I_{\alpha \in \left\{ L,R\right\} }\left( t\right) &=&-\int_{\alpha }d
\mathbf{r}\partial _{t}\rho \left( \mathbf{r},t\right) =-\sum_{k_{\alpha
}\in \alpha }\partial _{t}\sigma _{k_{\alpha }k_{\alpha }}\left( t\right)
\nonumber \\
&=&i\sum_{k_{\alpha }\in \alpha }\sum_{\ell \in D}\left( h_{D\alpha
,k_{\alpha }\ell }\sigma _{\alpha D,\ell k_{\alpha }}-\sigma _{D\alpha
,k_{\alpha }\ell }h_{\alpha D,\ell k_{\alpha }}\right)  \nonumber \\
&=&-tr\left[ Q_{\alpha }(t)\right]  \label{a5}
\end{eqnarray}

\subsection{Expressions of the dissipation function $Q_{\alpha}$ using the Green function formalism}
To calculate the dissipation term $Q_{\alpha }$ in EOM and transient current
equation, we use the time-dependent non-equilibrium Green function (TDNEGF) formalism.
We note that the overlap matrix $\mathbf{s}$ is treated as an identity matrix $\mathbf{I}$ when deriving the Green function formalism in the appendices.
This replacement of the overlap-matrix by the identity matrix has been verified to be valid if the hamiltonian matrix is modified according to mathematical techniques in the textbook \cite{book1}(Ch. 8.1.2), e.g. $\mathbf{h}_{D}-E\mathbf{s}_{D}=\mathbf{h}_{D}-E(\mathbf{s}_{D}-\mathbf{I})-E%
\mathbf{I}=\mathbf{h}_{D}^{\prime }-E\mathbf{I}$.
Appendix A demonstrates relevant approximations and derivations, and gives the formulae of $Q_{\alpha }$:
\begin{equation}
Q_{\alpha ,mn}(t)=-\sum_{\ell \in D}\int_{-\infty }^{\infty }d\tau \left[
G_{D,m\ell }^{<}\left( t,\tau \right) \Sigma _{\alpha ,\ell n}^{A}\left(
\tau ,t\right) +G_{D,m\ell }^{R}\left( t,\tau \right) \Sigma _{\alpha ,\ell
n}^{<}\left( \tau ,t\right) +H.c.\right]  \label{a7}
\end{equation}
We ignore the term associated with the complex-axis integral along the Keldysh contour $
\gamma _{K}$ (see Fig. \ref{fig_KC}) in Eq. (\ref{appxQ}). Green functions $\mathbf{G}
^{<}$ and $\mathbf{G}^{R}$ in Eq. (\ref{a7}) are calculated via Kadanoff-Baym equations
\cite{TDNEGF1,kb1} as derived by Eqs. (\ref{appxR}) and (\ref{appxL}) in appendix A:
\begin{eqnarray}
i\frac{d}{dt}\mathbf{G}_{D}^{R}\left( t,t^{\prime }\right) &=&\delta \left(
t-t^{\prime }\right) +\mathbf{h}_{D}\left( t\right) \mathbf{G}_{D}^{R}+
\mathbf{\Sigma }^{R}\cdot \mathbf{G}_{D}^{R}  \label{a8} \\
i\frac{d}{dt}\mathbf{G}_{D}^{<}\left( t,t^{\prime }\right) &=&\left[ \mathbf{
\Sigma }^{R}\cdot \mathbf{G}^{<}+\mathbf{\Sigma }^{<}\cdot \mathbf{G}^{A}
\right] \left( t,t^{\prime }\right) +\mathbf{h}\left( t\right) \mathbf{G}
^{<}\left( t,t^{\prime }\right)  \label{a9}
\end{eqnarray}
with notations $\left[ f\cdot g\right] \left( t,t^{\prime
}\right) =\int_{t_{0}}^{\infty }d\bar{t}f(t,\bar{t})g(\bar{t},t^{\prime })$
, $\mathbf{\Sigma }^{\lessgtr ,A,R}=\sum_{\alpha }\mathbf{\Sigma }_{\alpha
}^{\lessgtr ,A,R}$, and $f^{A}\left( t,t^{\prime }\right) =
\left[ f^{R}\left( t^{\prime },t\right) \right] ^{\dagger }$ \ $\left( f\in
G,\Sigma \right) $. The self-energy for electrodes by definition is given by
\begin{eqnarray}
\mathbf{\Sigma }_{\alpha }^{A}\left( t,t^{\prime }\right) &=&i\Theta \left(
t^{\prime }-t\right) \mathbf{h}_{D\alpha }(t)\exp \left\{
i\int_{t}^{t^{\prime }}\mathbf{h}_{\alpha }\left( \bar{t}\right) d\bar{t}
\right\} \mathbf{h}_{\alpha D}(t^{\prime })  \label{a10} \\
\mathbf{\Sigma }_{\alpha }^{<}\left( t,t^{\prime }\right) &=&i\mathbf{h}
_{D\alpha }(t)f_{\alpha }\left( \mathbf{h}_{\alpha ,t=t_{0}}\right) \exp
\left\{ i\int_{t}^{t^{\prime }}\mathbf{h}_{\alpha }\left( \bar{t}\right) d
\bar{t}\right\} \mathbf{h}_{\alpha D}(t^{\prime })  \label{a11}
\end{eqnarray}
Here, $\Theta \left( t^{\prime }-t\right) $ is the Heaviside step function, $
\mathbf{h}_{\alpha }$ is the Kohn-Sham matrix of the isolated electrode $\alpha$, and $
f_{\alpha }$ is the Fermi distribution function for $\alpha \in L,R$.

\subsection{Wide-band limit approximation for the dissipation function $Q_{\alpha}$}

For efficient computations of the equation of motion in Eqs. (\ref{a8}) and (\ref
{a9}), we introduce the wide-band limit (WBL) approximation \cite{WBL1}
for L and R electrodes by the following valid conditions \cite{TDNEGF1,WBL2}: (1) the bandwidths of the electrodes are
larger than the coupling strength between the device and L or R electrode; (2) the broadening matrix (the imaginary part of
self-energy for electrodes) is assumed to be energy-independent, resulting
in the requirement for an electrode's density of state and device-electrode
couplings to be slowly varying in energy; and (3) the level shifts of
electrodes via bias are approximated to be constant for all energy levels.

Solving the problem within the wide-band limit is not necessary. However WBL
considerably speeds up the calculation and is a very good approximation
model for simple metal contacts at comparatively low bias. The numerically approximated
self-energy is determined at the Fermi level of the systems without bias, and
is split up into two real matrices: one is the hermitian matrix $\Lambda
_{\alpha }$ for level shift, and the other is the anti-hermitian matrix $\Gamma _{\alpha }$ for level broadening. Equations. (\ref
{a10}) and (\ref{a11}) now can be rewritten as:
\begin{equation}
\mathbf{\Sigma }_{\alpha }^{R,A}\left( t,t^{\prime }\right) =\left( \mathbf{
\Lambda }_{\alpha }\mp i\mathbf{\Gamma }_{\alpha }\right) \delta \left(
t-t^{\prime }\right)   \label{a14}
\end{equation}
Here, $\mathbf{\Lambda }_{\alpha }$ and $\mathbf{\Gamma }_{\alpha }$ are
related by the Kramers-Kronig relation \cite{kk1}. According to the
derivation in appendix B, the dissipation term for electrodes L and R can be
given as \cite{thesis1}:
\begin{equation}
\mathbf{Q}_{\alpha }(t)=\mathbf{K}_{\alpha }(t)+\mathbf{K}_{\alpha
}^{\dagger }(t)+\left\{ \mathbf{\Gamma }_{\alpha },\mathbf{\sigma} \left( t\right)
\right\} +i\left[ \mathbf{\Lambda }_{\alpha },\mathbf{\sigma} \left( t\right) \right]
\label{a15}
\end{equation}
with the definition of $\mathbf{K}_{\alpha }(t)$ as:
\begin{eqnarray}
&&\mathbf{K}_{\alpha }(t)=-\frac{2i}{\pi }\mathbf{U}_{\alpha
}(t)\int_{-\infty }^{\infty }\frac{f_{\alpha }\left( \epsilon \right)
e^{i\epsilon t}}{\epsilon -\mathbf{h}_{D}(0)-\sum_{\alpha }\left( \mathbf{%
\Lambda }_{\alpha }-i\mathbf{\Gamma }_{\alpha }\right) }d\epsilon \mathbf{%
\Gamma }_{\alpha }  \label{a16} \\
&&-\frac{2i}{\pi }\int_{-\infty }^{\infty }\left[ \mathbf{I-U}_{\alpha
}(t)e^{i\epsilon t}\right] \frac{f_{\alpha }\left( \epsilon \right) }{%
\epsilon -\mathbf{h}_{D}(t)-\sum_{\alpha }\left( \mathbf{\Lambda }_{\alpha
}-i\mathbf{\Gamma }_{\alpha }\right) +V_{\alpha }\left( t\right) \mathbf{I}}%
d\epsilon \mathbf{\Gamma }_{\alpha }  \nonumber
\end{eqnarray}
and
\begin{equation}
\mathbf{U}_{\alpha }(t)=e^{-i\int_{0}^{t}\left[ \mathbf{h}_{D}(\bar{t}
)+\sum_{\alpha }\left( \mathbf{\Lambda }_{\alpha }-i\mathbf{\Gamma }_{\alpha
}\right) -V_{\alpha }\left( \bar{t}\right) \mathbf{I}\right] d\bar{t}}
\label{a16w}
\end{equation}
Together with the EOM for $\mathbf{\sigma }_{D}(t)$ in Eqs. (\ref{a3}) and (\ref{a4}), one now can calculate the transient electron density of the device and the boundary currents in Eq. (\ref{a5}).

\subsection{Calculations of self-energy matrices $\mathbf{\Lambda}$ and $\mathbf{\Gamma}$ }
We can express the retarded self-energy for the contact with electrode $\alpha$ in the energy domain \cite{thesis1} as:
\begin{equation}
\mathbf{\Sigma }_{\alpha }^{r}(E)=\mathbf{h}_{D\alpha } \mathbf{G}^{r}_{\alpha}(E) \mathbf{h}_{\alpha D}
\label{selfer}
\end{equation}
Considering the semi-infinite electrodes, the periodic Au(111) lattices can be divided into principle layers (PLs) along the transport direction (see Fig \ref{fig_TDNEGF}).
Here, we choose PLs to be wide enough so that only interactions between the nearest PLs need to be considered; i.e. the coupling matrix $\mathbf{h}_{D\alpha}$ between contact $\alpha$ and device region $D$ will be restricted to one PL. Consequently only the surface block of $\mathbf{G}^{r}_{\alpha}$, i.e. the surface green function $\mathbf{G}^{r,s}_{\alpha}$, is needed for calculating Eq. (\ref{selfer}). This work adopts an iterative method \cite{surfG} to calculate the surface green function that includes properties of the semi-infinite lattices. In principle, the self-energy matrices $\mathbf{\Gamma}$ and $\mathbf{\Lambda}$ for wide-band approximation are calculated at the Fermi level as
\begin{equation}
\mathbf{h}_{D\alpha } \mathbf{G}^{r,s}_{\alpha}(E_{F}) \mathbf{h}_{\alpha D}=\mathbf{\Lambda}_{\alpha}-i\mathbf{\Gamma}_{\alpha}
\label{selfer2}
\end{equation}

\subsection{Correction of the device Hamiltonian for transient variations of electron densities
using the capacitive network model}
For the open quantum system, the device Hamiltonian can take the perturbative
form \cite{hform1} of:
\begin{equation}
\mathbf{h}_{D}=\mathbf{h}_{D}^{0}\left( q_{0}\right) +\delta \mathbf{h}%
_{D}\left( \delta q\right)  \label{hd_t1}
\end{equation}%
Here, the change of electron density can be calculated by the density matrix
$\mathbf{\sigma }_{D}$ in Eq. (\ref{a3})
\begin{equation}
\delta n\left( \vec{r}\right) =\sum_{\mu \nu }Re[\rho _{\mu \nu }\chi _{\mu
}\left( \vec{r}\right) \chi _{\nu }^{\ast }\left( \vec{r}\right)
]-n_{0}\left( r\right)  \label{mullq1}
\end{equation}%
as a spatial distribution function, or, alternatively, by
\begin{equation}
\delta q_{i}=\sum_{\mu \in \{i\}}\sum_{\nu }Re[\rho _{\mu \nu }s_{D,\nu \mu
}]-q_{0,i}  \label{mullq2}
\end{equation}%
for atom $i$. Here, $n_{0}\left( r\right) $ and $q_{0,i}$ are the reference
atomic charges chosen for neutrality, $\mathbf{s}_{D}$ is the device overlap
matrix, and $\chi _{i}\left( \vec{r}\right) $ is a set of local basis
functions used in the tight-binding formulation. According to a Taylor expansion of the total energy around the reference density, this change of charge density
can result in corrections to the Hartree and the exchange-correlation potentials
\cite{perturb1,perturb2} for the device Hamiltonian as in Eq. (\ref{hd_t1}),
and it is continuously renewed with the density matrix in Eq. (\ref{a3}). In this work we simplify
the correction of the device Hamiltonian $\delta \mathbf{h}_{D}$ by
retaining only the Hartree potential $\delta V_{H}$ (assuming the
exchange-correlation term is insignificant in the mean-field scope), which obeys the three-dimensional Poisson
equation
\begin{equation}
\nabla ^{2}\delta V_{H}(r)=-\delta n\left( \vec{r}\right)  \label{Poisson}
\end{equation}%
with the boundary conditions imposed by the lead potentials. The conventional
Poisson solution is based on spatially-discretized grids for numerically iterative processes, and can be time-consuming for large systems. Thus, it is convenient to develop an
approximately analytical model.

Extending the idea of the muffin-tin (MT) approximation, each atom $i$ can
define a spherical region (MT-sphere) that bounds the total excess charges $%
\delta q_{i}$ from Eq. (\ref{mullq2}). The paired parts of charges inside different neighboring MT-spheres are considered as
capacitance effects \cite{datta1}. All these spheres of atoms in the system then further construct a capacitance network
architecture that supplies an analytical solution for the Poisson
equation \cite{blockbook1}. In principle,
the capacitances are treated as a
combination of the electrostatic capacitance $c_{e}$ and quantum capacitance
$c_{Q}$ \cite{datta1}. Herein, the quantum capacitance is assumed to be less
dominant than the electrostatic capacitance for $\delta q_{i}$ and
is ignored in our work.

Replacing the spatial solution ($\nabla _{r}^{2}$) of Poisson equation by
the atomic-site solution ($\nabla _{i}^{2}$) of the capacitive model
\cite{orth1,orth2}, we can rewrite Eq. (\ref{Poisson}) by a matrix-form
equation $\mathbf{\hat{C}\vec{V}}=\mathbf{\vec{Q}}$
\begin{eqnarray}
c_{ij} &=&4\pi \epsilon \frac{\bar{a}_{ij}^{2}}{|r_{ij}|}\left( 1+\frac{%
\bar{a}_{ij}^{2}}{|r_{ij}|^{2}-2\bar{a}_{ij}^{2}}+...\right)
\label{circuit1} \\
\hat{C}_{ij} &=&\sum_{k\in \{1NN\}_{i,con}}\delta _{i,j}c_{ik}+\sum_{k\in \{1NN\}_{i}}\delta _{i,j}c_{ik}-\delta
_{j,k}c_{ij}  \label{circuit2} \\
\vec{Q}_{i} &=&e\cdot \delta q_{i}+e\cdot \delta q_{d,i}+\sum_{k\in
\{1NN\}_{i,con.}}\delta _{j,k}c_{ij}V_{con.j}  \label{circuit3}
\end{eqnarray}%
Here, the matrix elements of $\mathbf{\hat{C}}$ are calculated in a two-center approximation as proposed in the tight-binding formulation \cite{thesis1}, obeying the formal condition
$e\delta q_{i}+e\delta q_{d,i}=\sum_{j}c_{ij}(\delta
V_{i}-\delta V_{j})$.
The notation $\{1NN\}_{i}$ is the group of the first nearest-neighbor (NN) atoms in the
device region for atom $i$, and $\{1NN\}_{i,con.}$ is the group of the first
nearest-neighbor atoms in the lead region for device atom $i$. Extending the solution by including more capacitively-coupling terms $\{nNN\}$ ($n\in 1,2,...$) is reasonable, because the additional capacitance terms $c_{ij}$ ($n\geq2$) diminish with the increasing separation $|r_{ij}|$ as indicated in Eq. (\ref{circuit1}). Herein, $c_{ij}$ defines
the capacitance between two ideal metal spheres, $|r_{ij}|$ is the spatial distance between atoms $i$ and $j$, and
$\bar{a}_{ij}$ is the effective muffin-tin radius for atoms $i$ and $j$ and is defined by $\bar{a}_{ij}=(r_{MT,i}+r_{MT,j})/4$ in this work.
$\mathbf{\vec{V}}\equiv (\delta V_{1},\delta V_{2},...,\delta V_{N})$ is the
potential vector with the components being the electrostatic potential for device atoms
$i\in \left\{ 1,...,N\right\} $. $V_{con.j}$ is the potential of lead atom $j
$ imposed by boundary conditions. $\delta q_{i}$ is the charge density
obtained by Eq. (\ref{mullq2}), and $\delta q_{d,i}$ represents the defect
charge for atom $i$. By linear algebra the potential vector $\mathbf{\vec{V}}$ can be easily
solved using $\mathbf{\vec{V}}=\mathbf{\hat{C}}^{-1}\mathbf{%
\vec{Q}}$.
For instance, in a 1-dimensional homogeneous system having 4
atoms L-A-A-R, the capacitance between nearby atoms is denoted as $c$, and the biases are denoted
as $v_{L}$ and $v_{R}$ for lead atoms L and R, respectively. There are no excess
charges ($\delta q=0$) inside the MT-sphere of device atoms A. In this way, the 2x2
capacitance matrix has components $\hat{C}_{11}=\hat{C}_{22}=2c$ and $\hat{C}%
_{12}=\hat{C}_{21}=-c$, and the charge vector is $\mathbf{\vec{Q}}^{t}=[
\begin{array}{cc}
cv_{L} & cv_{R}%
\end{array}] $. One then can obtain the electrostatic potentials for the two
atoms A as $\mathbf{\vec{V}}^{t}=[
\begin{array}{cc}
2v_{L}+v_{R} & v_{L}+2v_{R}%
\end{array}] /3$, which agree with the free-space Poisson solution.

\begin{figure}[tbp]
\begin{center}
\includegraphics[scale=0.55]{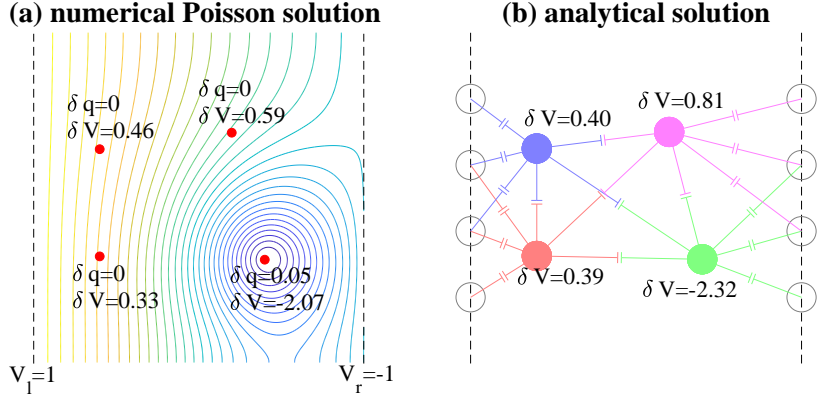}
\end{center}
\caption{Profile of Hartree potential $\delta V_{i}$ for the first structure with area $9\times 9$-$\AA^{2}$, solved by (a) the numerical Poisson solution and (b) the analytical solution. Four atoms with specified charges $\delta q$ (see Figure (a)) are placed between two leads and have $r_{MT}=1\AA$. In (a) the additional spatial function $\delta V(r)$ is illustrated by the contour curves. In (b) each boundary condition of the electrostatic potential is represented by four lead atoms.  } \label{poissonsolve1}
\end{figure}

\begin{figure}[tbp]
\begin{center}
\includegraphics[scale=0.55]{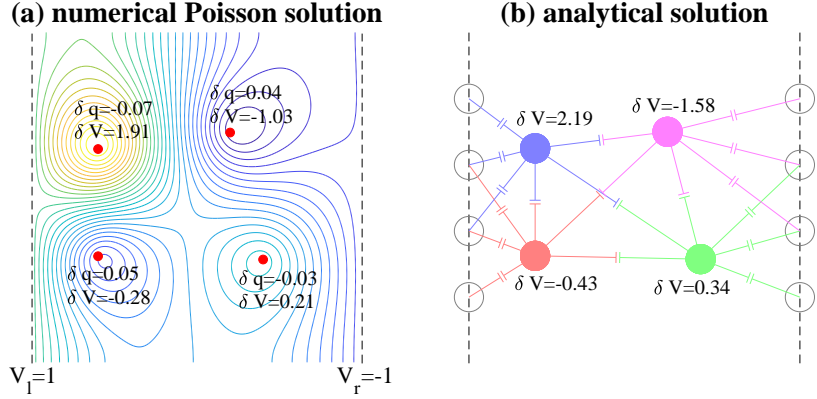}
\end{center}
\caption{Profile of Hartree potential $\delta V_{i}$ for the second structure with area $9\times 9$-$\AA^{2}$, solved by (a) the numerical Poisson solution and (b) the analytical solution. Relevant setups are the same with that in Fig. \ref{poissonsolve1}, except for atom charge $\delta q$.} \label{poissonsolve2}
\end{figure}

Figures (\ref{poissonsolve1}-\ref{poissonsolve2}) illustrate 2-dimensional
examples with a comparison between the numerically iterative solution and the analytical capacitance model. In order to solve the spatial Poisson equation in Eq. (\ref{Poisson}), the density function $\delta n_{i}\left( r\right)$
for two-dimensional systems is assumed to be:
\begin{equation}
\delta n_{i}\left( r\right) =\frac{ \delta q_{i}}{2 %
\pi \eta^{2}}e^{-\frac{|r-R_{i}|}{\eta}}  \label{disf}
\end{equation}
This is according to the symmetry assumption \cite{thesis1}, where $R_{i}$ is the position for atom $i$, and $\eta$ is associated with the effective radius of the MT-sphere by $\eta\propto r_{MT,i}$ (use $\eta= r_{MT,i}$ for examples in Figs. (\ref{poissonsolve1}-\ref{poissonsolve2})). The obtained potential $\delta V_{H}(r)$ is projected on the atomic sites through
\begin{equation}
\delta V_{i}=\frac{\int d\mathbf{r}\delta V_{H}(\mathbf{r})e^{-\frac{%
|r-R_{i}|}{\eta }}}{\int d\mathbf{r}e^{-\frac{|r-R_{i}|}{\eta }}}
\label{projr2a}
\end{equation}
for a comparison with the analytical solution $\mathbf{\vec{V}}$ in this work. We note that the analytical model associated with orientatingly capacitive couplings implies a spatial density function beyond the symmetry assumption. As indicated in Figs. (\ref{poissonsolve1}-\ref{poissonsolve2}), the analytical solution shows comparable results with that from the numerically-iterative method. Otherwise, it is emphasized that the analytical model turns inefficient at large biases or strong density variations, because the MT sphere cannot accurately account for the distorted and displaced electron density distribution from the nucleus.

The relevant parameters of the MT radius used herein are \cite{MTSi,MTO,MTAu,MTP} $r_{MT}(Si)=1.164{\AA}$, $r_{MT}(O)=0.947{\AA}$, $r_{MT}(Au)=1.376{\AA}$, and $r_{MT}(P)=1.377{\AA}$.
All computations are operated on a workstation having 2xCPU(E5-2690 v2) and 128G of DRAM. Fortran source codes can be downloaded online \citep{codeF}.

\section{Numerical results}
\subsection{Atomic Electrodes: Au(111) Nanotubes}
\begin{figure}[tbp]
\begin{center}
\includegraphics[scale=0.23]{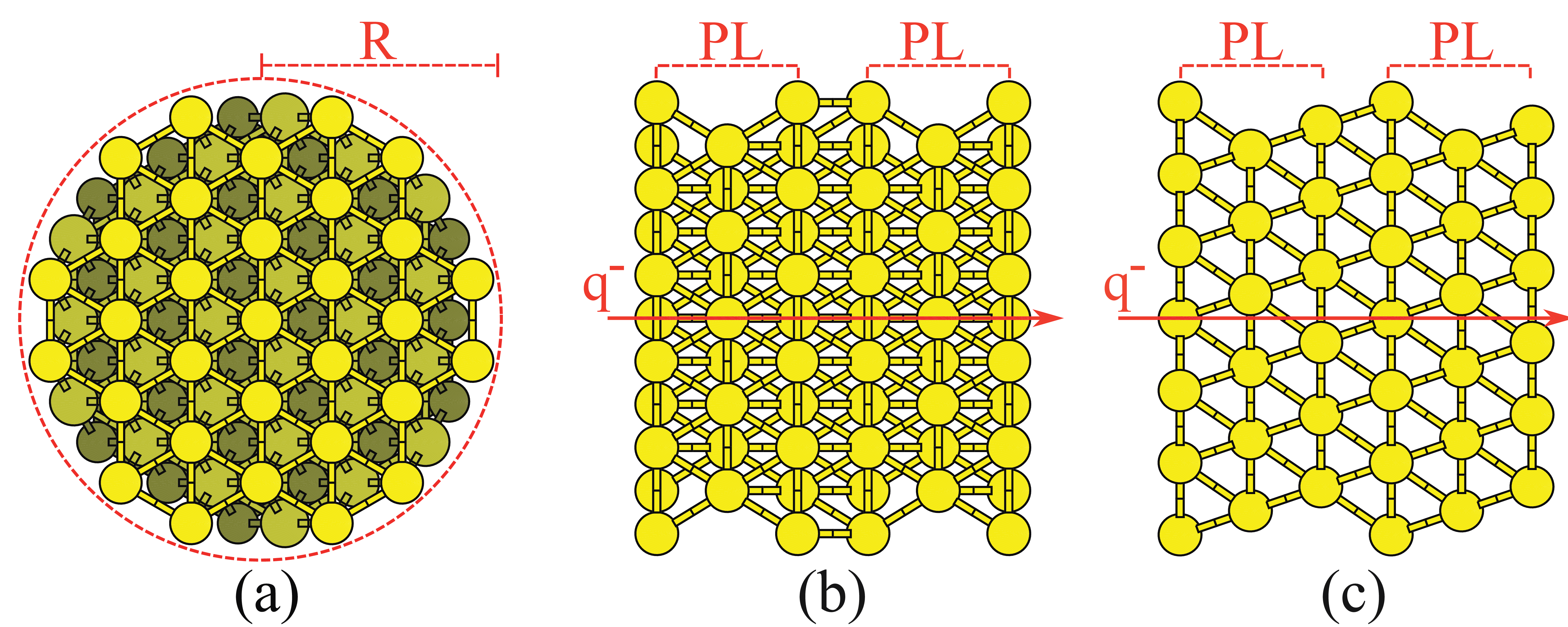}
\end{center}
\caption{Ball-stick representation of the Au(111) nanotube in (a) the longitudinal perspective and in (b-c) two lateral perspectives. The radius of the cross section in (a) is set as R=2a. Two of the principle layers (PL) arranged along the longitudinal (transport) direction in (b-c) represent a segment of the nanotube. $q^{-}$ indicates the quasi one dimensional (red arrow) charge transport.} \label{fig_au111}
\end{figure}
\begin{figure}[tbp]
\begin{center}
\includegraphics[scale=0.6]{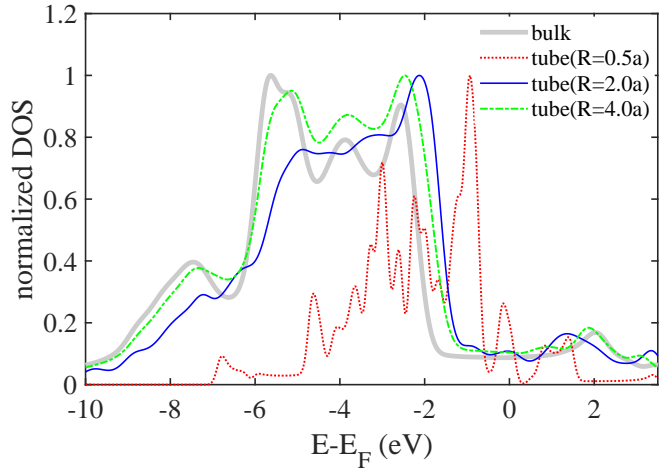}
\end{center}
\caption{Normalized density of states (DOS) for Au bulk and infinite Au(111) nanotubes, in which the radiuses of the nanotubes are set as R=0.5a, R=2.0a, and R=4.0a.} \label{fig_DOSau111}
\end{figure}

This research uses Au(111) nanotubes as atomic electrodes. The length $\ell$ of the Au-Au bond is determined with geometry relaxations of the Au bulk in the SIESTA program\cite{siesta1,siesta2}, and the obtained value is $\ell$=2.8785 ${\AA}$ (lattice constant a=$\sqrt{2}\ell$=4.0708 ${\AA}$, which is similar to the experimental value \cite{chembook1} of 4.0782 ${\AA}$). The effects of core electrons are evaluated with norm-conserving pseudopotentials in the local density approximation (Ceperley-Alder exchange-correlation potential\cite{LDA1,LDA2}), which are generated by the ATOM program\cite{atom1,siesta1}. The valence electrons of Au are calculated in the s-d hybridized configuration \cite{sdhybrid1}. All the calculations for nanotubes are performed on $8\times8\times8$ Monkhorst-Pack grids in reciprocal spaces under an electronic temperature of 300K. Figure \ref{fig_au111} shows (a) the longitudinal perspective and (b-c) two lateral perspectives for a finite segment of Au(111) nanotubes. In actual computations, the nanotube is set as an infinite stack of principle layers (PL) along the axial (longitudinal) direction, and has cross-section radius R. Figure \ref{fig_DOSau111} shows the normalized density of states (DOS) for Au bulk and Au(111) nanotubes, where the radiuses of the nanotubes are set as R=0.5a, R=2.0a, and R=4.0a, respectively. $E_{F}$ is the Fermi level corresponding to the mentioned system. In Fig. \ref{fig_DOSau111}, DOS of the Au bulk shows metallic properties as the literature \cite{AuBulkDOS} reports. When increasing the cross-section radius R, the DOS functions of Au(111) nanotubes at energies near $E_{F}$ change from discrete to uniform distributions, depicting the transfer of systems from 1D-nanotube to 3D-bulk structures. In this work, we use Au(111) nanotubes with R=2a for semi-infinite electrodes in transport problems. This adoption (setting R=2a) meets the requirement of slowly-varying DOS for the wide-band limit (WBL) condition \cite{WBL1}, and demands computation resources that are affordable.

\subsection{Doped Si-SiO$_{2}$ quantum dots}

\begin{figure}[tbp]
\begin{center}
\includegraphics[scale=0.25]{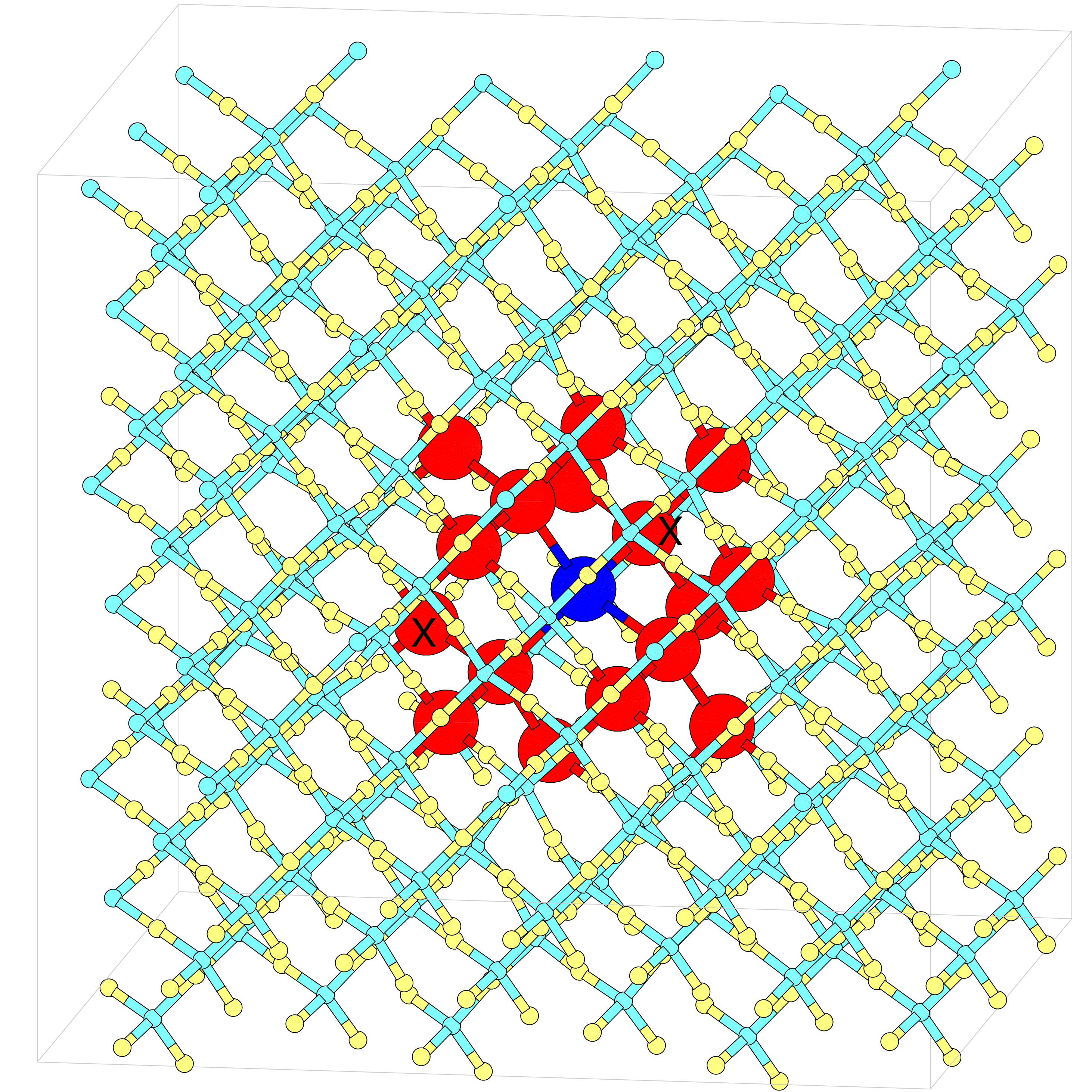}
\end{center}
\caption{Schematics of Si Quantum Dot (red atoms) embedded in SiO$_{2}$ matrix (light cyan-yellow atoms). The phosphorus atom P (blue atom) is doped inside the quantum dot for the 1P-doping condition. Two red atoms with X marks denote the doping locations inside the quantum dot and at the interface, respectively, for the 2P-doping condition.} \label{fig_cell}
\end{figure}

\begin{figure}[tbp]
\begin{center}
\includegraphics[scale=0.8]{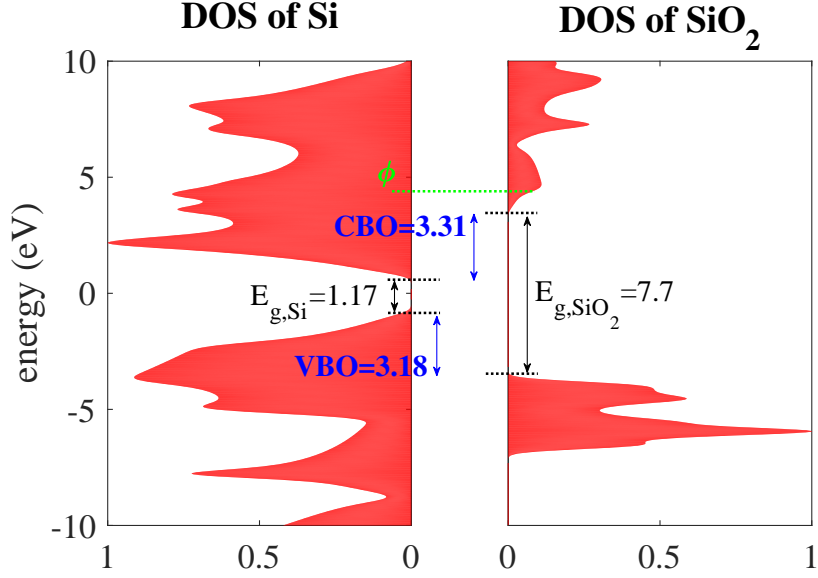}
\end{center}
\caption{Band diagrams of Si-SiO$_{2}$-slabs heterojunction by Anderson's rule. The density of state at equilibrium is arranged according to a hypothetical flat vacuum level.
The computed energy gaps are $E_{g,Si}$=1.17 eV and $E_{g,SiO_{2}}$=7.7 eV. The valence band offset VBO is 3.18 eV and the conduction band offset CBO is 3.31 eV.} \label{fig_DOSsisio2}
\end{figure}

This research investigates the silicon quantum dots with diameters around 1.0 nm that are embedded in a $\beta$-cristobalite SiO$_{2}$ matrix. The dopant phosphorus (P) atoms are placed inside quantum dots due to their energetically-favored formation of structures \cite{QD1} (see Fig. \ref{fig_cell}). Lattice constants are determined with geometry relaxations in the SIESTA program (set orbital bases s and p for species Si, O, and P). The obtained values are 5.5001 $\AA$ (5.4306 by experiment \cite{chembook1}) for the Si diamond structure and 7.46831 $\AA$ (7.160-7.403 $\AA$ in textbooks \cite{chembook1,sio2a}) for  $\beta$-cristobalite silica. We investigate the energy band diagram of Si-SiO$_2$-slabs heterojunctions by using Anderson's rule through Fig. \ref{fig_DOSsisio2}, in which the vacuum levels (green dotted lines) of Si and SiO$_2$ slabs are aligned at the same energy. Here, the vacuum level is defined as the effective potential $\phi$ (adding local pseudopotential, Hartree potential, and exchange-correlation potential) at zero-density points near the surface of slabs having 35 atomic layers.  All calculations are performed at $\Gamma$-point of the reciprocal space. As indicated in Fig. \ref{fig_DOSsisio2}, the vacuum levels are 1.064 eV and 1.626 eV for Si-slab and SiO$_{2}$-slab, respectively, corresponding to working functions $W_{Si}$=4.46 eV and $W_{SiO_{2}}$=4.52 eV. The experimental value \cite{chembook1} is $4.60\leq W_{Si}\leq 4.91$ eV. The computed energy gaps are 1.17 eV for bulk silicon and 7.7 eV for $\beta$-cristobalite silica, which can be compared with the experimental values of 1.1 eV and 9.0 eV, respectively. The valence band offset (VBO) and conduction band offset (CBO) for Si-SiO$_{2}$ heterojunctions are estimated to be 3.18 eV and 3.31 eV, respectively. The obtained VBO values are smaller than experimental measurements \cite{vbo1,vbo2} with VBO=4.6 eV and CBO=3.1 eV. Several theoretical works using hopping mechanisms \cite{QD1,QD2,QD3} give VBO$\approx$2.6 eV and CBO$\approx$3.9 eV.

\begin{figure}[tbp]
\begin{center}
\includegraphics[scale=0.6]{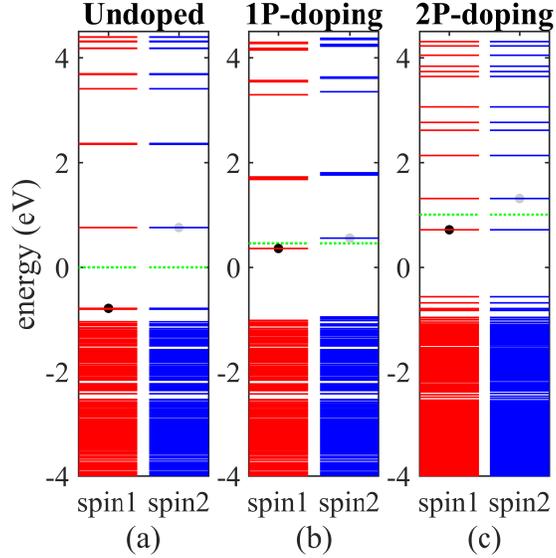}
\end{center}
\caption{Spin-up and spin-down spectra of (a) undoped, (b) 1P-doping,and (c) 2P-doping systems. Energies are aligned using the embedding SiO$_2$ states, and are shifted with the reference of the fermi level of the undoped structure. Black and gray circles mark HOMO and LUMO states, respectively. The green dotted line represents the fermi level of the corresponding system.} \label{fig_eigval}
\end{figure}

With relevant material parameters, the Si-SiO$_{2}$ quantum-dot device in Fig. \ref{fig_cell} is constructed from a $3\times3\times3$ supercell of $\beta$-cristobalite silica by removing O atoms in a cut-off box. Figure \ref{fig_eigval} reports the eigenvalue spectra for the undoped, 1P-doping, and 2P-doping structures after relaxation processes, using the corresponding initial geometries in Fig. \ref{fig_cell}. The spectrum energies are aligned using the SiO$_2$ states (deep valence states), and the energy axes show the common origin according to the fermi level of the undoped structure. Black and gray circles mark the highest occupied molecular orbital (HOMO) and lowest unoccupied molecular orbital (LUMO) states, respectively. The green dotted line represents the fermi level of the corresponding structure. In Fig. \ref{fig_eigval}(a), the undoped structure exhibits a distinguished energy spectrum from that of the slab-heterojunction in Fig. \ref{fig_DOSsisio2}, revealing the essential mechanism for strain-induced electron levels \cite{mismatch1}. For the 1P-doping system, the odd number of electrons leads to the spin-dependent energy spectrum in Fig. \ref{fig_eigval}(b), which depicts a clear donor behavior and agrees well with previous works \cite{QD1,donor1}. This study adopts the 2P-dopping structure in Fig. \ref{fig_eigval}(c) due to the following considerations: (i) high conductivity at a low bias V owing to the rising fermi level and the decreasing energy gap, compared to the undoped structure; and (ii) having spin independence for better computational efficiency and the negligible spin-flip mechanism, compared to the 1P-doping case.

\subsection{Steady (time-independent) electron transport in open quantum-dot systems}
\begin{figure}[tbp]
\begin{center}
\includegraphics[scale=0.275]{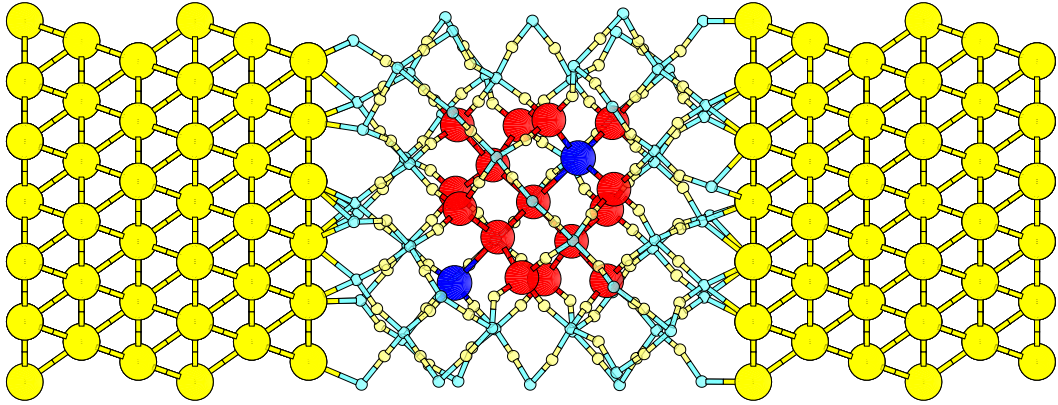}
\end{center}
\caption{Schematics of a Si-based (red spheres) quantum dot embedded in $\beta$-cristobalite SiO$_{2}$ matrix (small light cyan-yellow spheres), where two dopant atoms (blue phosphorus spheres) are placed inside the quantum dot and at the Si-SiO$_2$ interface, respectively. The device is enclosed between two semi-infinitely-long Au(111) wires (larger yellow spheres) with an applied voltage. } \label{fig_qd_dev}
\end{figure}
Fig. \ref{fig_qd_dev} constructs the open transport system of quantum dots. The device region, i.e. Si-based quantum dot (red spheres) and SiO$_{2}$ matrix (small light cyan-yellow spheres), is enclosed by two semi-infinitely long Au wires. Two dopant atoms (phosphorus; blue spheres) are placed inside the quantum dot and at the Si-SiO$_2$ interface, respectively, according to their energetically-favored formation energy\cite{QD1}. It is assumed that the positions of atoms of Au electrodes are under constraint by the experimental set-ups, while the atoms of the doped Si-SiO$_{2}$ quantum dot are in equilibrium according to geometry relaxations by the SIESTA program. The distance between the nearest cross sections of silica and gold boundaries before geometry relaxations is initially set to be 1.8 $\AA$ in this work.

\begin{figure}[tbp]
\begin{center}
\includegraphics[scale=0.65]{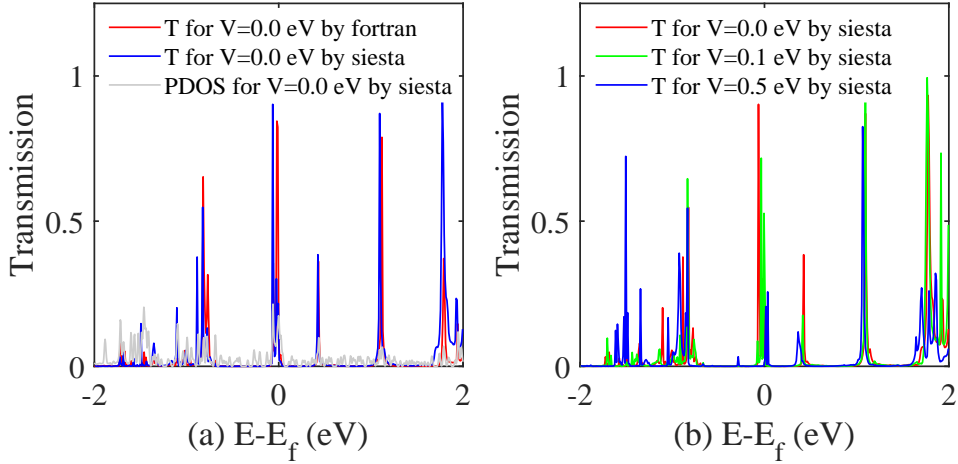}
\end{center}
\caption{(a) Transmission functions of the quantum-dot system (V=0.0 eV) using SIESTA and the fortran programs, to be compared with the projected density of state (PDOS) of Si-SiO$_{2}$ QDs. (b) Transmission functions of the quantum-dot system calculated by SIESTA programs at different voltage biases.  } \label{fig_T}
\end{figure}

By applying non-equilibrium green functions for steady transport problems \cite{datta1}, the transmission function of the quantum-dot system is obtained as shown in Fig. \ref{fig_T}. In Fig. \ref{fig_T}(a), the transmission function T (blue curve) is calculated by SIESTA::Transiesta programs, and is compared with the projected density of state (PDOS; gray curve) of the Si-SiO$_2$ quantum dot. The red-curve transmission function is calculated by the fortran program which extracts the relevant hamiltonian and overlap matrices from SIESTA for modeling the tight-binding formulation, and the numerical framework is employed in time-dependent non-equilibrium green functions for transient problems below. Numerical results demonstrate that (i) the conductance channels in the transmission function T are associated with the density of states of the Si-SiO$_{2}$ quantum dots as expected, and (ii) the tight-binding formulation works well since the comparable transmission functions by SIESTA and fortran programs. Figure \ref{fig_T}(b) depicts transmission functions for several systems calculated by SIESTA programs with different bias setups. It is found that the transmission profiles of the devices non-linearly vary with biases, and reveal considerable effects of charge fluctuations inside the device. It is emphasized that SIESTA is the standard time-independent density functional program without the mentioned approximations for this work. We also note that the quantum-dot system presents non-zero conductance at near zero bias, which is similar to the analysis in Nuria's work \cite{QD1}. This conductance associated with the finite transmission function at the fermi level, however, decreases with an increasing bias as shown in Fig. \ref{fig_T}(b).

\subsection{Transient (time-dependent) electron transport in open quantum-dot systems}
This section studies the time-dependent electron transport for the quantum-dot system in Fig. \ref{fig_qd_dev}. Additional parameters and numerical methods are as follows: time step $\delta t=5as$, voltage function $V_{f}(t)=V_{dc}\left[1-exp^{-t/\tau}\right]+V_{ac}cos(\omega t)$ with $\tau=2fs$, globally-adaptive integrator treating singularities in the energy domain, and the fourth-order Runge Kutta methods (RK4) for solving the time-differential equation. Here, we adopt the linear extrapolation of the density matrix $\sigma_{D}$ (Eq. \ref{a3}) during the iterative process by RK4.

\begin{figure}[tbp]
\begin{center}
\includegraphics[scale=0.6]{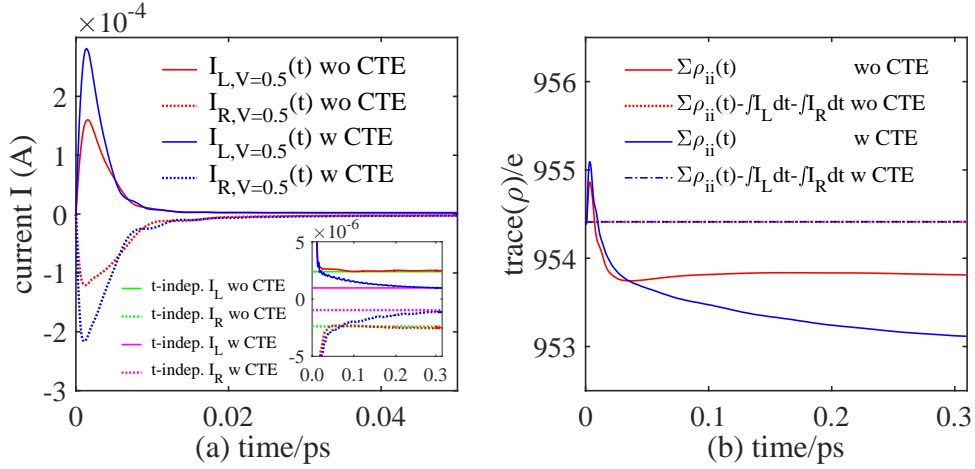}
\end{center}
\caption{(a) Transient current I of quantum-dot devices at a symmetry DC bias $V_{dc}$=0.5V with/without including charge transfer effect (CTE). The inset diagram shows that the currents asymptotically approach the value of the steady solution by SIESTA (green curves). (b) Transient charge numbers of the quantum-dot device with/without including CTE. After applying bias, a part of the electrons ($\Sigma\rho_{ii}$, solid curves) participate in the inter-orbital transferring process ($\int I_{L}dt+\int I_{R}dt$). The dot curves show the conservation of total charges.} \label{fig_TD1}
\end{figure}

Figure \ref{fig_TD1}(a) shows transient currents by our TDNEGF codes with and without corrections for charge transfer effects (CTE). The voltage functions are symmetrically set by $V_{L}=V_{f,V_{dc}=0.5V,V_{ac}=0V}$ and $V_{R}=V_{f,V_{dc}=-0.5V,V_{ac}=0V}$ for the left and right electrodes, respectively. Moreover, the time-independent solutions for steady currents are derived via Landauer Buttiker formula \cite{datta1}, an integral of the transmission functions in Fig. \ref{fig_T}(b). We note that the integrals using the transmission function for $V=0.0$ eV in Fig. \ref{fig_T}(b) correspond to the steady current without CTE, and that using $V=0.5$ eV gives the current with CTE. With the comparison between the steady and transient results in Figure \ref{fig_TD1}(a), We observe that the transient currents asymptotically approach the values of the corresponding steady solutions \cite{longtime1} by SIESTA (see the inset diagram), no matter the charge transfer effects are considered or not. This concludes the validation of the TDNEGF program as well as the proposed analytical model for treating CTE. Moreover, the inclusion of charge transfer effects presents considerable corrections for the convergence of the transient current, suggesting non-trivial variations/excitations of charge in the device. The curves also depict that the calculation including CTE requires a much longer time to bring the system into the steady sate, inferring a slow redistribution process of the charge density. Figure \ref{fig_TD1}(b) shows the transient electron number of the device and the integrals of boundary currents, obeying the continuity equation for the device region.

\begin{figure}[tbp]
\begin{center}
\includegraphics[scale=0.575]{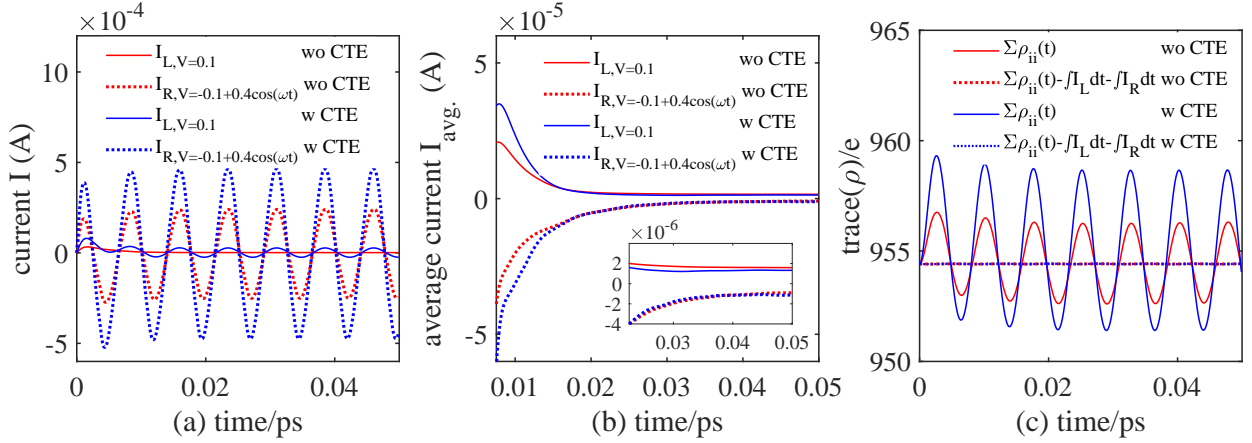}
\end{center}
\caption{(a) Transient current I of quantum-dot devices at AC bias $V_{L}$=0.1V and $V_{R}$=-0.1+0.4$cos(\omega t)$ V, with/without including a charge transfer effect (CTE). Here the AC frequency is $\omega=0.835\times 10^{15}$. By including CTE, $I_{L}$ exhibits continuous oscillations of interfacial currents. (b) Effective currents $I_{avg.}$ by averaging $I$(t) in (a) over one period $T=2\pi/\omega$. The inset diagram shows that the effective currents, with/without including CTE, asymptotically converge into similar values. (c) Transient charge numbers of the quantum-dot device with and without including CTE. The dot curves show the conservation of total charges.} \label{fig_TD2}
\end{figure}

In Fig. \ref{fig_TD2}, we study the transient currents for the quantum-dot devices under AC bias voltages. To observe the diffusion of CTE through the device, the voltage functions are asymmetrically set by $V_{L}=V_{f,V_{dc}=0.1V,V_{ac}=0V}$ and $V_{R}=V_{f,V_{dc}=-0.1V,V_{ac}=0.4V}$ for the left and right electrodes, respectively. Here the AC frequency is $\omega=0.835\times 10^{15}$. In Fig. \ref{fig_TD2}(a), for the quantum-dot device with charge transfer effects, the interfacial current $I_{L}$ of the left electrode exhibits continuous oscillations; while $I_{L}$ quickly declines to a steady value for the case without charge transfer effects. The interfacial current $I_{R}$ of the right electrode, however, is always oscillatory due to the driving of the AC potentials at the local (right) electrode. This observation identifies the AC-induced oscillation of charge densities inside the quantum-dot device via CTE algorithms. To correlate with the alternative DC measurements, we calculate the average current $I_{avg.}(t)$ by averaging $I(t)$ over one period $T=2\pi/\omega$. Numerical results of $I_{avg.}(t)$ are shown in Fig. \ref{fig_TD2}(b). In its inset diagram, the effective currents, no matter with and without charge transfer effects, asymptotically converge into similar values. We attribute the similarity of the asymptotical currents to the low DC bias, which contributes insignificant charge fluctuations on average, as suggested by the alike curves for V=0.0 and 0.1 eV in Fig. \ref{fig_T}(b). Figure \ref{fig_TD2}(c) validates the algorithms by the continuity equation in AC cases.

\section{Conclusions}
This research proposes an approximate analytical model to efficiently calculate the transient properties of quantum-dot systems under time-dependent external potentials.
Numerical results in the low DC bias and long-time limits present good agreements with the corresponding steady solutions, no matter the charge transfer effects are included or excluded.
For the cases using asymmetric AC biases, numerical calculations for transient currents also show distinct characteristics between the systems with and without charge transfer effects, revealing the essential oscillations/excitations of charge densities inside the device.

\section{Acknowledgement}
This work was supported by ChiMei Visual Technology Corporation.

\appendix
\section{Equation of motion for Green functions}
The Hamiltonian operator $\mathbf{\hat{h}}$ for open transport
systems without spin notations can be given by%
\begin{equation}
\mathbf{\hat{h}}=\sum_{k_{\alpha }}h_{\alpha ,k_{\alpha }}n_{k_{\alpha
}}+\sum_{m,n}h_{D,mn}a_{m}^{\dagger }a_{n}+\sum_{m,k_{\alpha }}h_{D\alpha
,mk_{\alpha }}a_{m}^{\dagger }a_{k_{\alpha }}+h_{\alpha D,k_{\alpha
}m}a_{k_{\alpha }}^{\dagger }a_{m}  \label{Hsys}
\end{equation}
The first term describes the $\alpha _{th}$ electrode with state $k_{\alpha }$,
the second term is for the device in geometry region D, and the third term is
for the coupling between the device and the electrode $\alpha $.

In this appendix, the algorithm of the time-dependent non-equilibrium green function is addressed for systems under the following conditions:
during $t<t_{0}$, the system is in thermal equilibrium at an inverse
temperature $\beta $ and chemical potential $\mu $. At times $t\geq t_{0}$, the system departs from the equilibrium conditions after applying external voltages.
Here, the response of the spin to external fields is ignored, resulting in diagonal green-function and self-energy matrices with respect to the spin parameter. The initial condition is defined by the ground state of the system.

\begin{figure}[tbp]
\begin{center}
\includegraphics[scale=0.5]{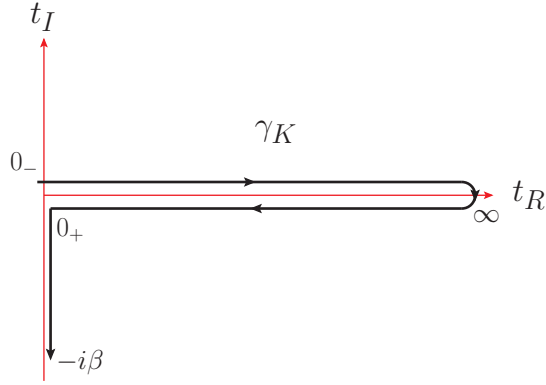}
\end{center}
\caption{The Keldysh contour $\gamma_{K}$ is an oriented contour having endpoints at $0_{-}$ and $-i\beta$. $\beta$ is the inverse temperature.
The contour is composed of a forward branch going from $t=0_{-}$ to t=$\infty$, a backward branch coming back from $t=\infty$ to $t=0{+}$, and a vertical (thermic) track
on the imaginary times axis between $0_{+}$ and $-i\beta$. $z$ and $z^{\prime}$ define variables along $\gamma_{K}$.} \label{fig_KC}
\end{figure}

For time-dependent electron transport problems, one begins with the one-particle green function on the Keldysh contour $\gamma _{K}$ (see Fig. \ref{fig_KC}). This green
function is defined as the ensemble average of the contour-ordered product of electron creation and annihilation operators in the Heisenberg picture,
\begin{equation}
G_{rs}\left( z,z^{\prime }\right) =-i\left\langle T_{C}\left[ a_{r}\left(
z\right) a_{s}^{\dagger }\left( z^{\prime }\right) \right] \right\rangle
\label{gdef}
\end{equation}
Here, $r,s$ present states of L(left-electrode), R(right-electrode), and D(device). $z$ and $z^{\prime }$ define complex variables along the contour $\gamma _{K}$. $T_{C}$ is the time-ordering operator. The creation operator $a^{\dagger }$ and annihilation operator $a$\ obey the equation of motion%
\begin{eqnarray}
i\frac{d}{dz}a\left( z\right) &=&\left[ a\left( z\right) ,h\left( z\right) %
\right] =h\left( z\right) a\left( z\right)  \label{eq1} \\
i\frac{d}{dz}a^{\dagger }\left( z\right) &=&\left[ a^{\dagger }\left(
z\right) ,h\left( z\right) \right] =-h\left( z\right) a^{\dagger }\left(
z\right)  \label{eq2}
\end{eqnarray}
Here, the anticommutator relation for fermions have been used, i.e. $\left\{
a_{r},a_{s}\right\} =\left\{ a_{r}^{\dagger },a_{s}^{\dagger }\right\} =0$
and $\left\{ a_{r},a_{s}^{\dagger }\right\} =\left\langle r|s\right\rangle $
with orthonormal state bases $r$ and $s$. Combining Eqs. (\ref{eq1}-\ref{eq2})
and (\ref{gdef}), the equation of motion for the green function can be given as:
\begin{eqnarray}
\left[ i\frac{d}{dz}-h\left( z\right) \right] G\left( z,z^{\prime }\right)
&=&\delta \left( z,z^{\prime }\right) 1  \label{h1} \\
G\left( z,z^{\prime }\right) \left[ -i\frac{d}{dz^{\prime }}-h\left(
z^{\prime }\right) \right] &=&\delta \left( z,z^{\prime }\right) 1
\label{h2}
\end{eqnarray}
Here, the green function follows Kubo-Martin-Schwinger (KMS) boundary
conditions on the imaginary axis in $\gamma _{K}$. $h\left( z\right) $ is
the single-particle Hamiltonian.

Applying the definition in Eq. (\ref{a2}), the matrix structure of $\mathbf{G}$ has block matrix form:%
\begin{equation}
G\mathbf{=}\left[
\begin{array}{cccc}
\mathbf{G}_{L} & \mathbf{G}_{LD} & \mathbf{G}_{LR} \\
\mathbf{G}_{DL} & \mathbf{G}_{D} & \mathbf{G}_{DR} \\
\mathbf{G}_{RL} & \mathbf{G}_{RD} & \mathbf{G}_{R}%
\end{array}%
\right]  \label{hm}
\end{equation}
Equation (\ref{h1}) in matrix form is hence given by
\begin{equation}
i\frac{d}{dz}\mathbf{G}\left( z,z^{\prime }\right) -\mathbf{h}\left(
z\right) \mathbf{G}\left( z,z^{\prime }\right) =\delta \left( z,z^{\prime
}\right) \mathbf{1}  \label{eqgf1}
\end{equation}%
Here, the equations for components $\mathbf{G}_{\alpha D}$ and $\mathbf{G}_{D}$ are
\begin{eqnarray}
\left[ i\frac{d}{dz}-\mathbf{h}_{D}\left( z\right) \right] \mathbf{G}
_{D}\left( z,z^{\prime }\right) &=&\delta \left( z,z^{\prime }\right)
\mathbf{1}+\sum_{\alpha \in L,R}\mathbf{h}_{D\alpha }\left( z\right)
\mathbf{G}_{\alpha D}\left( z,z^{\prime }\right)  \label{eqtmp3} \\
\left[ i\frac{d}{dz}-\mathbf{h}_{\alpha }\left( z\right) \right] \mathbf{G}
_{\alpha D}\left( z,z^{\prime }\right) &=&\mathbf{h}_{\alpha D}\left(
z\right) \mathbf{G}_{D}\left( z,z^{\prime }\right)  \label{eqtmp4}
\end{eqnarray}
By multiplying Eq. (\ref{eqtmp4}) with the green function $\mathbf{G}
_{\alpha }$, i.e. $\left[ -i\frac{d}{dz^{\prime }}-\mathbf{h}_{\alpha
}\left( z^{\prime }\right) \right] \mathbf{G}_{\alpha }\left( z,z^{\prime
}\right) =\delta \left( z,z^{\prime }\right) \mathbf{1}$ in Eq. (\ref{h1}),
we can obtain $\mathbf{G}_{\alpha D}\left( z,z^{\prime }\right) $ as
\begin{eqnarray*}
\int_{\gamma _{K}}d\bar{z}\mathbf{G}_{\alpha }\left( z,\bar{z}\right) \left[
i\frac{d}{d\bar{z}}-\mathbf{h}_{\alpha }\left( \bar{z}\right) \right]
\mathbf{G}_{\alpha D}\left( \bar{z},z^{\prime }\right) &=&\int_{\gamma _{K}}d
\bar{z}\left[ \left( -i\frac{d}{d\bar{z}}-\mathbf{h}_{\alpha }\left( \bar{z}
\right) \right) \mathbf{G}_{\alpha }\left( z,\bar{z}\right) \right] \mathbf{G
}_{\alpha D}\left( \bar{z},z^{\prime }\right) \\
&=&\int_{\gamma _{K}}d\bar{z}\delta \left( z,\bar{z}\right) \mathbf{G}
_{\alpha D}\left( \bar{z},z^{\prime }\right) \\
&=&\int_{\gamma _{K}}d\bar{z}\mathbf{G}_{\alpha }\left( z,\bar{z}\right)
\mathbf{h}_{\alpha D}\left( \bar{z}\right) \mathbf{G}_{D}\left( \bar{z}
,z^{\prime }\right)
\end{eqnarray*}
\begin{equation}
\mathbf{G}_{\alpha D}\left( z,z^{\prime }\right) =\int_{\gamma _{K}}d\bar{z}
\mathbf{G}_{\alpha }\left( z,\bar{z}\right) \mathbf{h}_{\alpha D}\left( \bar{
z}\right) \mathbf{G}_{D}\left( \bar{z},z^{\prime }\right)  \label{glead1}
\end{equation}
We apply integration by parts and assume the disappearance of electrons
at infinite distance. Inserting equation (\ref{glead1}) into
equation (\ref{eqtmp3}), the equation of motion for $\mathbf{G}_{D}\left(
z,z^{\prime }\right) $ can be obtained as
\begin{eqnarray*}
\left[ i\frac{d}{dz}-\mathbf{h}_{D}\left( z\right) \right] \mathbf{G}
_{D}\left( z,z^{\prime }\right) &=&\delta \left( z,z^{\prime }\right)
\mathbf{1} \\
&&+\int_{\gamma _{K}}d\bar{z}\left[ \sum_{\alpha }\mathbf{h}_{D\alpha
}\left( z\right) \mathbf{G}_{\alpha }\left( z,\bar{z}\right) \mathbf{h}
_{\alpha D}\left( \bar{z}\right) \right] \mathbf{G}_{D}\left( \bar{z}
,z^{\prime }\right)
\end{eqnarray*}
The term $\sum_{\alpha }\mathbf{h}_{D\alpha }\left( z\right) \mathbf{G}
_{\alpha }\left( z,\bar{z}\right) \mathbf{h}_{\alpha D}\left( \bar{z}\right)
=\sum_{\alpha }\mathbf{\Sigma }_{\alpha }$ is defined as the coupling
self-energy $\mathbf{\Sigma }\left( z,\bar{z}\right) $ and the equation is reformulated as
\begin{equation}
\left[ i\frac{d}{dz}-\mathbf{h}_{D}\left( z\right) \right] \mathbf{G}
_{D}\left( z,z^{\prime }\right) =\delta \left( z,z^{\prime }\right) \mathbf{1
}+\int_{\gamma _{K}}d\bar{z}\mathbf{\Sigma }\left( z,\bar{z}\right) \mathbf{G
}_{MM}\left( \bar{z},z^{\prime }\right)  \label{eqmm}
\end{equation}

\subsection{Kadanoff-Baym equations}
The equations for the device's green function $\mathbf{G}_{D}\left( z,z^{\prime
}\right) $ are summarized as:
\begin{eqnarray}
\left[ i\frac{d}{dz}-\mathbf{h}\left( z\right) \right] \mathbf{G}\left(
z,z^{\prime }\right) &=&\delta \left( z,z^{\prime }\right) \mathbf{1}
+\int_{\gamma _{K}}d\bar{z}\mathbf{\Sigma }\left( z,\bar{z}\right) \mathbf{G}
\left( \bar{z},z^{\prime }\right)  \label{b1} \\
\mathbf{G}\left( z,z^{\prime }\right) \left[ -i\frac{d}{dz^{\prime }}-
\mathbf{h}\left( z^{\prime }\right) \right] &=&\delta \left( z,z^{\prime
}\right) \mathbf{1}+\int_{\gamma _{K}}d\bar{z}\mathbf{G}\left( z,\bar{z}
\right) \mathbf{\Sigma }\left( \bar{z},z^{\prime }\right)  \label{b2}
\end{eqnarray}
Here, the subscript $D$ is dropped in this subsection for simplicity. Because
the lesser green function $\mathbf{G}^{<}$ is directly related to
observable physical quantities, i.e. electron densities and currents, its
integro-differential equation is described first. Using the definition of $
\mathbf{G}^{<}\left( t_{-},t_{+}\right) =\mathbf{G}\left( z=t_{-},z^{\prime
}=t_{+}\right) $ with $t_{-}<t_{+}$ and separating the
Keldysh contour by real and imaginary segments in Eq. (\ref{b1}), one gets:
\begin{eqnarray}
&&\left[ i\frac{d}{dt_{-}}\right] \mathbf{G}^{<}\left( t_{-},t_{+}\right) -
\mathbf{h}\left( t_{-}\right) \mathbf{G}^{<}\left( t_{-},t_{+}\right)
\nonumber \\
&=&\int_{Re\gamma _{K}}d\bar{t}\left[ \mathbf{\Sigma }\left( t_{-},\bar{t}
\right) \mathbf{G}\left( \bar{t},t_{+}\right) \right] -i\int_{Im\gamma
_{K}}d\tau \left[ \mathbf{\Sigma }\left( t_{-},t_{0}-i\tau \right) \mathbf{G}
\left( t_{0}-i\tau ,t_{+}\right) \right]  \label{b4}
\end{eqnarray}
Adopting common notations $f$ for green functions $\mathbf{G}$ and self-energy $\mathbf{\Sigma}$ in the
Keldysh space, we arrive at:
\begin{eqnarray}
f\left( t,t^{\prime }\right) |_{f\in \mathbf{G,\Sigma }} &=&f^{\delta
}\left( t\right) \delta \left( t-t^{\prime }\right) +\Theta \left(
t-t^{\prime }\right) f^{>}\left( t,t^{\prime }\right) +\Theta \left(
t^{\prime }-t\right) f^{<}\left( t,t^{\prime }\right)  \label{aa1} \\
f^{R}\left( t,t^{\prime }\right) |_{f\in \mathbf{G,\Sigma }} &=&f^{R,\delta
}\left( t\right) \delta \left( t-t^{\prime }\right) +\Theta \left(
t-t^{\prime }\right) \left[ f^{>}\left( t,t^{\prime }\right) -f^{<}\left(
t,t^{\prime }\right) \right]  \label{aa2} \\
f^{A}\left( t,t^{\prime }\right) |_{f\in \mathbf{G,\Sigma }} &=&f^{A,\delta
}\left( t\right) \delta \left( t-t^{\prime }\right) -\Theta \left( t^{\prime
}-t\right) \left[ f^{>}\left( t,t^{\prime }\right) -f^{<}\left( t,t^{\prime
}\right) \right]  \label{aa3} \\
f^{\rceil }\left( t,\tau \right) |_{f\in \mathbf{G,\Sigma }} &=&f^{<}\left(
t,t_{0}-i\tau \right)  \label{aa4} \\
f^{\lceil }\left( \tau ,t\right) |_{f\in \mathbf{G,\Sigma }} &=&f^{>}\left(
t_{0}-i\tau ,t\right)  \label{aa5}
\end{eqnarray}
Equation (\ref{b4}) can be rewritten as
\begin{eqnarray*}
&&i\frac{d}{dt_{-}}\mathbf{G}^{<}\left( t_{-},t_{+}\right) -\mathbf{h}\left(
t_{-}\right) \mathbf{G}^{<}\left( t_{-},t_{+}\right) \\
&=&\int_{t_{0}}^{\infty }d\bar{t}\mathbf{\Sigma }^{R}\left( t_{-},\bar{t}
\right) \mathbf{G}^{<}\left( \bar{t},t_{+}\right) +\int_{t_{0}}^{\infty }d
\bar{t}\mathbf{\Sigma }^{<}\left( t_{-},\bar{t}\right) \mathbf{G}^{A}\left(
\bar{t},t_{+}\right) -i\int_{0}^{\beta }d\tau \mathbf{\Sigma }^{\rceil
}\left( t_{-},\tau \right) \mathbf{G}^{\lceil }\left( \tau ,t_{+}\right)
\end{eqnarray*}
Alternatively, we have
\begin{equation}
i\frac{d}{dt_{-}}\mathbf{G}^{<}\left( t_{-},t_{+}\right) -\mathbf{h}\left(
t_{-}\right) \mathbf{G}^{<}\left( t_{-},t_{+}\right) =\left[ \mathbf{\Sigma }
^{R}\cdot \mathbf{G}^{<}+\mathbf{\Sigma }^{<}\cdot \mathbf{G}^{A}+\mathbf{%
\Sigma }^{\rceil }\star \mathbf{G}^{\lceil }\right] \left( t_{+},t_{-}\right)
\label{b5}
\end{equation}
with notations $\left[ f\cdot g\right] \left( t,t^{\prime
}\right) =\int_{t_{0}}^{\infty }d\bar{t}f(t,\bar{t})g(\bar{t},t^{\prime })$
and $\left[ f\star g\right] \left( t,t^{\prime }\right) =-i\int_{0}^{\beta }d
\bar{t}f(t,\tau )g(\tau ,t^{\prime })$. Equations for the greater green
function $\mathbf{G}^{>}$ can be obtained by similar processes:
\begin{eqnarray}
i\frac{d}{dt_{-}}\mathbf{G}^{<}\left( t_{-},t_{+}\right) -\mathbf{h}\left(
t_{-}\right) \mathbf{G}^{<}\left( t_{-},t_{+}\right) &=&\left[ \mathbf{
\Sigma }^{R}\cdot \mathbf{G}^{<}+\mathbf{\Sigma }^{<}\cdot \mathbf{G}^{A}+
\mathbf{\Sigma }^{\rceil }\star \mathbf{G}^{\lceil }\right] \left(
t_{-},t_{+}\right)  \label{appx1} \\
i\frac{d}{dt_{+}}\mathbf{G}^{>}\left( t_{+},t_{-}\right) -\mathbf{h}\left(
t_{+}\right) \mathbf{G}^{>}\left( t_{+},t_{-}\right) &=&\left[ \mathbf{
\Sigma }^{R}\cdot \mathbf{G}^{>}+\mathbf{\Sigma }^{>}\cdot \mathbf{G}^{A}+
\mathbf{\Sigma }^{\rceil }\star \mathbf{G}^{\lceil }\right] \left(
t_{+},t_{-}\right)  \label{appx2}
\end{eqnarray}
and,
\begin{eqnarray}
-i\frac{d}{dt_{+}}\mathbf{G}^{<}\left( t_{-},t_{+}\right) -\mathbf{h}\left(
t_{+}\right) \mathbf{G}^{<}\left( t_{-},t_{+}\right) &=&\left[ \mathbf{G}%
^{R}\cdot \mathbf{\Sigma }^{<}+\mathbf{G}^{<}\cdot \mathbf{\Sigma }^{A}+%
\mathbf{G}^{\rceil }\star \mathbf{\Sigma }^{\lceil }\right] \left(
t_{-},t_{+}\right)  \label{appx3} \\
-i\frac{d}{dt_{-}}\mathbf{G}^{>}\left( t_{+},t_{-}\right) -\mathbf{h}\left(
t_{-}\right) \mathbf{G}^{>}\left( t_{+},t_{-}\right) &=&\left[ \mathbf{G}%
^{R}\cdot \mathbf{\Sigma }^{>}+\mathbf{G}^{>}\cdot \mathbf{\Sigma }^{A}+%
\mathbf{G}^{\rceil }\star \mathbf{\Sigma }^{\lceil }\right] \left(
t_{+},t_{-}\right)  \label{appx4}
\end{eqnarray}%
Equations (\ref{appx1})-(\ref{appx4}) are the Kadanoff-Baym equations with
symmetry relations of functions $f\in \mathbf{G,\Sigma }$:
\begin{eqnarray}
f^{\gtrless }\left( t,t^{\prime }\right) |_{f\in \mathbf{G,\Sigma }} &=&-
\left[ f^{\gtrless }\left( t^{\prime },t\right) \right] ^{\dagger }
\label{b6} \\
f^{\rceil \lceil }\left( t,t^{\prime }\right) |_{f\in \mathbf{G,\Sigma }}
&=&-\left[ f^{\rceil \lceil }\left( t^{\prime },t\right) \right] ^{\dagger }
\label{b7} \\
G^{>}\left( t,t\right) &=&-i+G^{<}\left( t,t\right) ,\text{ at equal time}
\label{b8} \\
G^{A}\left( t,t^{\prime }\right) &=&\left[ G^{R}\left( t^{\prime },t\right)
\right] ^{\dagger }
\end{eqnarray}

\subsection{Approximate equations for fast numerical implementation by neglecting the complex-axis integral}
For the equation of motion for the retarded green function $\mathbf{G}^{R}\left(
t,t^{\prime }\right) $, one can differentiate Eq. (\ref{aa2})
with respect to $t$, ignoring the $\mathbf{G}^{\delta }\delta \left(
t-t^{\prime }\right) $ term and the complex path in the Keldysh contour,
\begin{eqnarray}
i\frac{d}{dt}\mathbf{G}^{R}\left( t,t^{\prime }\right) &=&i\delta \left(
t-t^{\prime }\right) \left[ \mathbf{G}^{>}\left( t,t^{\prime }\right) -
\mathbf{G}^{<}\left( t,t^{\prime }\right) \right]  \nonumber \\
&&+\Theta \left( t-t^{\prime }\right) \left[ i\frac{d}{dt}\mathbf{G}
^{>}\left( t,t^{\prime }\right) -i\frac{d}{dt}\mathbf{G}^{<}\left(
t,t^{\prime }\right) \right]  \label{b9}
\end{eqnarray}
Together with Eqs. (\ref{appx1}) and (\ref{appx2}), Eq. (\ref{b9}) can be
rewritten as
\begin{eqnarray}
i\frac{d}{dt}\mathbf{G}^{R}\left( t,t^{\prime }\right) &=&\delta \left(
t-t^{\prime }\right) +\mathbf{h}\left( t\right) \Theta \left( t-t^{\prime
}\right) \left[ \mathbf{G}^{>}\left( t,t^{\prime }\right) -\mathbf{G}
^{<}\left( t,t^{\prime }\right) \right]  \nonumber \\
&&+\Theta \left( t-t^{\prime }\right) \left[ \mathbf{\Sigma }^{R}\cdot
\mathbf{G}^{>}+\mathbf{\Sigma }^{>}\cdot \mathbf{G}^{A}-\mathbf{\Sigma }
^{R}\cdot \mathbf{G}^{<}-\mathbf{\Sigma }^{<}\cdot \mathbf{G}^{A}\right]
\left( t,t^{\prime }\right)  \nonumber \\
&=&\delta \left( t-t^{\prime }\right) +\mathbf{h}\left( t\right) \mathbf{G}
^{R}+\mathbf{\Sigma }^{R}\cdot \mathbf{G}^{R}  \label{appxR}
\end{eqnarray}

For the equation of motion for the lesser green function $\mathbf{G}^{<}\left(
t,t^{\prime }\right) $, Eq. (\ref{appx1}), by ignoring complex integration,
gives
\begin{equation}
i\frac{d}{dt}\mathbf{G}^{<}\left( t,t^{\prime }\right) -\mathbf{h}\left(
t\right) \mathbf{G}^{<}\left( t,t^{\prime }\right) =\left[ \mathbf{\Sigma }%
^{R}\cdot \mathbf{G}^{<}+\mathbf{\Sigma }^{<}\cdot \mathbf{G}^{A}\right]
\left( t,t^{\prime }\right)  \label{appxL}
\end{equation}

For the evaluation of the dissipation term $\mathbf{Q}_{\alpha }$ in Eq. (\ref{a4}), we calculate the equation of motion for the lesser green function $
\mathbf{\sigma }\left( t\right) =-i\mathbf{G}^{<}\left( t,t\right) $ in the
equal-time limit. Substituting Eq. (\ref{b2}) from Eq. (\ref{b1}) and
applying the limit condition $t_{-}\approx t_{+}$\ during similar
derivations of Eq. (\ref{b5}), one obtains
\begin{eqnarray}
i\frac{d}{dt}\mathbf{G}^{<}\left( t,t\right) -\left[ \mathbf{h}\left(
t\right) ,\mathbf{G}^{<}\left( t,t\right) \right] &=&\left[ \mathbf{\Sigma }
^{R}\cdot \mathbf{G}^{<}+\mathbf{\Sigma }^{<}\cdot \mathbf{G}^{A}-\mathbf{G}
^{R}\cdot \mathbf{\Sigma }^{<}-\mathbf{G}^{<}\cdot \mathbf{\Sigma }^{A}
\right] \left( t,t\right)  \nonumber \\
&=&-\left[ \mathbf{G}^{R}\cdot \mathbf{\Sigma }^{<}+\mathbf{G}^{<}\cdot
\mathbf{\Sigma }^{A}\right] \left( t,t\right) +h.c.  \label{appxE}
\end{eqnarray}
Here, the complex integration has been ignored and the relations in Eq. (\ref
{b6}) are used. By comparing Eq. (\ref{appxE}) with Eq. (\ref{a3}) and using
$\mathbf{\sigma }\left( t\right) =-i\mathbf{G}^{<}\left( t,t\right) $, the
dissipation term can be given by
\begin{equation}
\mathbf{Q=}-\left[ \mathbf{G}^{R}\cdot \mathbf{\Sigma }^{<}+\mathbf{G}
^{<}\cdot \mathbf{\Sigma }^{A}\right] \left( t,t\right) +h.c.  \label{appxQ}
\end{equation}

\section{Wide-Band Limit approximation for the dissipation term $Q_{\alpha }$}

By applying the assumptions of the wide-band limit approximation, the advanced self-energy for
L and R in Eq. (\ref{a10}) becomes
\begin{eqnarray}
\Sigma _{\alpha ,mn}^{A}\left( t,t^{\prime }\right) &=&i\Theta \left(
t^{\prime }-t\right) \sum_{k_{\alpha }}h_{D\alpha ,mk_{\alpha }}(t)\exp
\left\{ i\int_{t}^{t^{\prime }}\epsilon _{k_{\alpha }}+V_{\alpha }\left(
\bar{t}\right) d\bar{t}\right\} h_{\alpha D,k_{\alpha }n}(t^{\prime })
\nonumber \\
&\simeq &\Theta \left( t^{\prime }-t\right) \int_{-\infty }^{\infty
}d\epsilon e^{i\epsilon \left( t^{\prime }-t\right) }\left[ i\cdot
h_{D\alpha ,m\overline{k}_{\alpha }}(t)e\exp \left\{ i\int_{t}^{t^{\prime
}}V_{\alpha }\left( \bar{t}\right) d\bar{t}\right\} h_{\alpha D,\overline{k}_{\alpha
}n}(t^{\prime })\right]  \nonumber \\
&=&\delta \left( t^{\prime }-t\right) \left[ \Lambda _{\alpha ,mn}+i\Gamma
_{\alpha ,mn}\right]  \label{bb1}
\end{eqnarray}
where the matrix in the square bracket in the last line is approximated by
the initial $\mathbf{\Sigma }_{\alpha }^{A}\left( \epsilon _{F}\right) $ at
fermi level of the unbiased system \cite{thesis1}. $V_{\alpha }(t)$ is the
external potential that is turned on at $t>t_{0}$, resulting in
time-dependent level shifts of $\alpha \in \{L,R\}$. The summation over all
single-electron states in the electrodes is replaced by an integration over
the entire energy, i.e. $\sum_{k_{\alpha }}\rightarrow \int_{-\infty
}^{\infty }d\epsilon $. The retarded/advanced self-energies are $\mathbf{%
\Sigma } _{\alpha }^{R,A}\left( t,t^{\prime }\right) =\left[ \Lambda
_{\alpha ,mn}\mp i\Gamma _{\alpha ,mn}\right] \delta \left( t^{\prime
}-t\right) $.

The lesser self-energy in Eq. (\ref{a11}) is
\begin{eqnarray}
\Sigma _{\alpha ,mn}^{<}\left( t,t^{\prime }\right)  &=&\sum_{k_{\alpha
}}h_{D\alpha ,mk_{\alpha }}(t)G_{\alpha ,k_{\alpha }}^{<}h_{\alpha
D,k_{\alpha }m}(t^{\prime })  \nonumber \\
&=&\sum_{k_{\alpha }}h_{D\alpha ,mk_{\alpha }}(t)h_{\alpha D,k_{\alpha
}m}(t^{\prime })\left[ i\cdot f_{\alpha }\left( \epsilon _{k_{\alpha
}}\right) e^{i\epsilon \left( t^{\prime }-t\right) }e^{i\int_{t}^{t^{\prime
}}V_{\alpha }\left( \bar{t}\right) d\bar{t}}\right]   \nonumber \\
&=&\frac{2i}{\pi }\Gamma _{\alpha ,mn}e^{i\int_{t}^{t^{\prime }}V_{\alpha
}\left( \bar{t}\right) d\bar{t}}\int_{-\infty }^{\infty }f_{\alpha }\left(
\epsilon \right) e^{i\epsilon \left( t^{\prime }-t\right) }d\epsilon
\label{bb2}
\end{eqnarray}%
From Eq. (\ref{a8}), the lesser green function can be solved as
\begin{equation}
\mathbf{G}_{D}^{R}\left( t,t^{\prime }\right) =-i\Theta \left( t-t^{\prime
}\right) e^{-i\int_{0}^{t}\left[ \mathbf{h}_{D}(\bar{t})+\sum_{\alpha
}\left( \mathbf{\Lambda }_{\alpha }-i\mathbf{\Gamma }_{\alpha }\right) %
\right] d\bar{t}}e^{-i\int_{t^{\prime }}^{0}\left[ \mathbf{h}_{D}(\bar{t}%
)+\sum_{\alpha }\left( \mathbf{\Lambda }_{\alpha }-i\mathbf{\Gamma }_{\alpha
}\right) \right] d\bar{t}}  \label{bb3}
\end{equation}%
Inserting Eqs. (\ref{bb1})-(\ref{bb3}) into Eq. (\ref{a7}), the dissipation term
for electrodes L and R can be given by
\begin{eqnarray}
\mathbf{Q}_{\alpha }(t) &=&-\int_{-\infty }^{\infty }d\tau \left[ \mathbf{G}%
_{D}^{<}\left( t,\tau \right) \mathbf{\Sigma }_{\alpha }^{A}\left( \tau
,t\right) +\mathbf{G}_{D}^{R}\left( t,\tau \right) \mathbf{\Sigma }_{\alpha
}^{<}\left( \tau ,t\right) +H.c.\right]   \nonumber \\
&=&-\int_{-\infty }^{\infty }d\tau \left[ \mathbf{G}_{D}^{R}\left( t,\tau
\right) \mathbf{\Sigma }_{\alpha }^{<}\left( \tau ,t\right) +H.c.\right]
+\left\{ \mathbf{\Gamma }_{\alpha },\sigma \left( t\right) \right\} +i\left[
\mathbf{\Lambda }_{\alpha },\sigma \left( t\right) \right]   \label{bb5}
\end{eqnarray}%
Here, the first term of the integration in the second line can be calculated
by
\begin{eqnarray}
&&\mathbf{K}_{\alpha }(t)=-\int_{-\infty }^{\infty }d\tau \left[ \mathbf{G}%
_{D}^{R}\left( t,\tau \right) \mathbf{\Sigma }_{\alpha }^{<}\left( \tau
,t\right) \right]   \nonumber \\
&=&\frac{-2}{\pi }\int_{-\infty }^{0}d\tau e^{-i\int_{\tau }^{t}\left[
\mathbf{h}_{D}(\bar{t})+\sum_{\alpha }\left( \mathbf{\Lambda }_{\alpha }-i%
\mathbf{\Gamma }_{\alpha }\right) -V_{\alpha }\left( \bar{t}\right) \right] d%
\bar{t}}\int_{-\infty }^{\infty }f_{\alpha }\left( \epsilon \right)
e^{i\epsilon \left( t-\tau \right) }d\epsilon \mathbf{\Gamma }_{\alpha }
\nonumber \\
&&+\frac{-2\Theta \left( t-\tau \right) }{\pi }\int_{0}^{\infty }d\tau
e^{-i\int_{\tau }^{t}\left[ \mathbf{h}_{D}(\bar{t})+\sum_{\alpha }\left(
\mathbf{\Lambda }_{\alpha }-i\mathbf{\Gamma }_{\alpha }\right) -V_{\alpha
}\left( \bar{t}\right) \right] d\bar{t}}\int_{-\infty }^{\infty }f_{\alpha
}\left( \epsilon \right) e^{i\epsilon \left( t-\tau \right) }d\epsilon
\mathbf{\Gamma }_{\alpha }  \nonumber \\
&=&\frac{-2i}{\pi }\mathbf{U}_{\alpha }(t)\int_{-\infty }^{\infty }\frac{%
f_{\alpha }\left( \epsilon \right) e^{i\epsilon t}}{\epsilon -\mathbf{h}%
_{D}(0)-\sum_{\alpha }\left( \mathbf{\Lambda }_{\alpha }-i\mathbf{\Gamma }%
_{\alpha }\right) }d\epsilon \mathbf{\Gamma }_{\alpha }  \nonumber \\
&&-\frac{2i}{\pi }\int_{-\infty }^{\infty }\left[ \mathbf{I-U}_{\alpha
}(t)e^{i\epsilon t}\right] \frac{f_{\alpha }\left( \epsilon \right) }{%
\epsilon -\mathbf{h}_{D}(t)-\sum_{\alpha }\left( \mathbf{\Lambda }_{\alpha
}-i\mathbf{\Gamma }_{\alpha }\right) +V_{\alpha }\left( t\right) \mathbf{I}}%
d\epsilon \mathbf{\Gamma }_{\alpha }  \label{bb6}
\end{eqnarray}%
with%
\begin{equation}
\mathbf{U}_{\alpha }(t)=e^{-i\int_{0}^{t}\left[ \mathbf{h}_{D}(\bar{t}%
)+\sum_{\alpha }\left( \mathbf{\Lambda }_{\alpha }-i\mathbf{\Gamma }_{\alpha
}\right) -V_{\alpha }\left( \bar{t}\right) \mathbf{I}\right] d\bar{t}}
\label{bb6w}
\end{equation}%
Conclusively, the dissipation term now is
\begin{equation}
\mathbf{Q}_{\alpha }(t)=\mathbf{K}_{\alpha }(t)+\mathbf{K}_{\alpha
}^{\dagger }(t)+\left\{ \mathbf{\Gamma} _{\alpha },\mathbf{\sigma} \left( t\right) \right\} +i%
\left[ \mathbf{\Lambda} _{\alpha },\mathbf{\sigma} \left( t\right) \right]   \label{bb7}
\end{equation}%
with the definition of $\mathbf{K}_{\alpha }(t)$ in Eq. (\ref{bb6}).

\end{document}